# Dust Formation by Bubble-burst Phenomenon at the Surface of a Liquid Steel Bath

A. G. GUÉZENNEC, J. C. HUBER,[1] F. PATISSON, Ph. SESSIECQ, J. P. BIRAT[1] and D. ABLITZER

LSG2M, UMR 7584 CNRS-INPL, Ecole des Mines, Parc de Saurupt, 54042 Nancy Cedex, France.
E-mail: guezenne@mines.inpl-nancy.fr          1) IRSID, Voie Romaine, BP 30320, 57283 Maizières-lès-Metz Cedex, France.

We have developed an experimental device for studying the main mechanism of dust formation in electric arc furnace steelmaking: the burst of gas bubbles at the liquid steel surface. As in the case of the air–water system, the bubble-burst process takes place in three steps: breaking of the film cap, projection of film drops, and projection of jet drops. The film break and the jet drop formation are observed with a high-speed video camera. The film drop aerosol enters a particle counter, which characterizes the drops in size and number. Results are presented and discussed. The quantification of both types of projections leads to the conclusion that the film drop projections represent the major source of dust. The amount of film drops greatly decreases with the parent bubble size. Bubbles with diameter under 4 mm theoretically do not produce film drops. Decreasing the CO-bubble size enough would therefore represent an effective solution for reducing drastically the electric arc furnace dust emission.

## 1. Introduction

The Electric Arc Furnace (EAF), designed for steelmaking from recycled ferrous scrap, co-produces between 15 and 25 kg of dust per ton of steel. This dust consists of metal oxides, lime, and silica, and contains zinc, lead and cadmium. These hazardous, leachable elements require EAF dust to be stored in specific landfills.

Therefore, the management of dust accounts for a significant part of the EAF running costs which is likely to increase in the coming years. Indeed, steel production by melting scrap in EAF goes on developing, whereas the fraction of zinc-coated materials in scrap is continuously growing up. Moreover, the legislation ruling the management, recovery and recycling of the industrial wastes is becoming more and more demanding. These aspects, combined with the growing concern about environmental issues, led steelmakers to imagine a policy of EAF dust management based on two objectives: reducing the amount of dust produced, and enhancing dust value thanks to the recovery of the zinc it contains.

In order to define the best operating conditions to achieve this strategy, it is necessary to understand and quantify the phenomena involved in EAF dust formation. In this paper, we first report the mechanisms of EAF dust formation. We then focus on the study of the major source of emission: the burst of CO bubbles on the steel bath surface. We present some theoretical information and the results obtained with the original experimental apparatus we developed at LSG2M.

## 2. The Mechanisms of Dust Formation in EAF

The dust collected in bag filters at the end of the EAF fume extraction system is the final product of a series of phenomena such as the emission of particles from the steel bath, the transport of these particles by the gas flow in the fume extraction system, the in-flight physico-chemical transformations they undergo, *etc*. The size of the particles contained in the EAF dust is smaller than 20 $\mu$m in diameter; 80% of them are below 2 $\mu$m.[1] We distinguish here two steps in the dust formation process: first, the emission of dust "precursors", *i.e.* vapors, metal droplets, and solid particles, inside the furnace; second, the conversion of those precursors into dust by agglomeration and physico-chemical transformations.

### 2.1. The Formation of Dust Precursors

A first study carried out by Huber[1] has clarified the emission mechanisms of dust precursors (see **Fig. 1**):
- volatilization, especially prominent at the hot spots in the arc zone (1) and the oxygen jet zone (1′), but taking place as well in the CO bubbles;
- emission of droplets at the impact points of the arc (2) and of the oxygen jet (2′) on the steel bath;
- projection of fine droplets by bursting of CO bubbles (3) coming from the decarburization of the steel bath;
- bursting of droplets (4) in contact with an oxidizing atmosphere within the surface;[2,3] the occurrence of this phenomenon, which can be classed as a bubble-burst mechanism, is uncertain in EAF;
- direct fly-off of solid particles (5) during the introduction of powder materials into the EAF (scrap, coal for slag

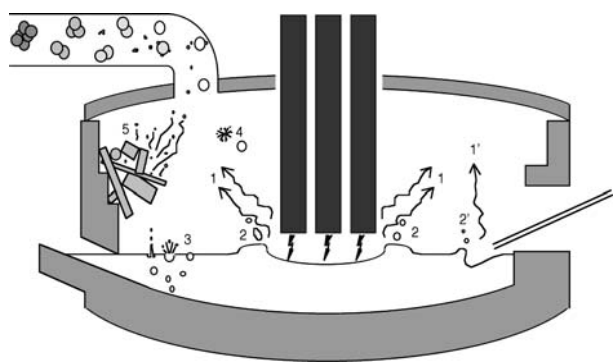

**Fig. 1.** Schematic representation of the mechanisms of dust formation in EAF.

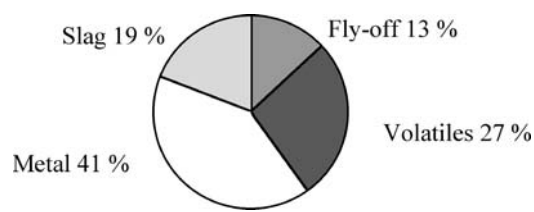

**Fig. 2.** Part of metal and slag projections, volatilization and direct fly-off in final EAF dust.[4]

foaming, additions, recycled dust, *etc*.).

The contribution of each mechanism to the formation of dust has been determined from an experimental study using tracers in a pilot furnace at IRSID.[4] **Figure 2** shows the repartition in weight of different materials in the final dust: droplets of liquid steel and liquid slag, volatile species contained in the liquid bath like zinc, particles coming from direct fly-off.

From these results and the preceding analysis, the prevailing mechanisms for dust precursor emission appear to be volatilization (27% of the dust) and bursting of CO bubbles (60% of the dust). The direct fly-off of solid particles remains very limited if sufficient operating precautions are taken. Concerning the projections at the impact points of the arc and of the oxygen jet, the size of the corresponding ejected drops varies between a tenth and a few millimeters.[1] Since no particles of such a size are found in EAF dust, it is likely that these projections fail to be carried up by the fume extraction system and fall back down in the liquid bath.

### 2.2. The Transformation of Precursors into Dust

The precursors are further transformed during their transport within the furnace and then in the fume extraction system. They can undergo physical transformations: condensation of the vapors, rapid solidification of the fine projections in contact with a colder atmosphere, in-flight agglomeration and coalescence of dust particles. The precursors can also be modified by chemical reactions (*e.g.* oxidation) with the carrier gas, whose temperature and composition vary, and possibly, they can react with other precursor particles. For a reaction between condensed phases (liquid or solid) to occur, particles must first be brought into contact. Therefore, there is a strong link between the mechanisms of agglomeration and the chemical evolution. These phenomena were already studied by Huber *et al*.[5] and will not be further dealt with here.

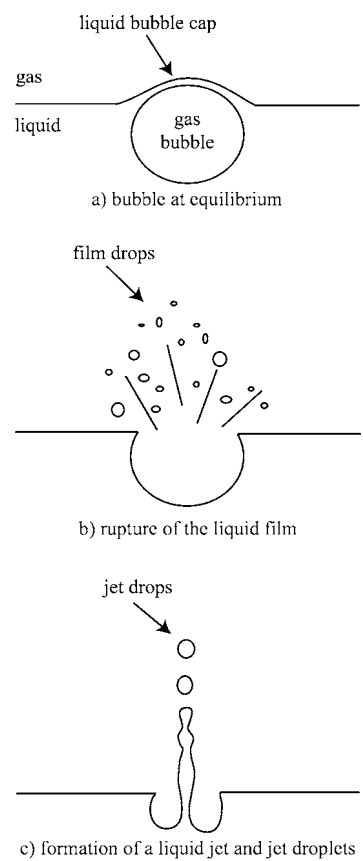

**Fig. 3.** Schematic representation of the burst of a bubble on a liquid surface.

### 2.3. The Phenomenon of Bubble-burst at a Liquid Surface

The projection of liquid steel and slag droplets by bursting of CO bubbles has been recognized as the principal mechanism of dust emission in EAF. Very few studies concerning bubble-burst at the surface of liquid metal have been reported.[6] However, in order to understand the phenomenon, useful results and observations can be found in the abundant literature about the air-water system.[7–17] From these studies, the bubble-burst process can be split up into three steps which give rise to two types of droplets (**Fig. 3**).

When emerging at the surface (Fig. 3(a)), a bubble lifts up a liquid film that progressively gets thinner under the influence of drainage, when the bubble comes to rest. The equilibrium position of a bubble floating at the surface of a liquid can be determined by following the approach proposed by Unger *et al*.[16]

As the film reaches a critical thickness, it ruptures and the bubble cap disintegrates into fine droplets called film drops (Fig. 3(b)). There is still some controversy about the process of rupture of the liquid film covering the bubble and the formation of the film drops. Indeed, the duration of the phenomenon (a few hundredths of microsecond), the small size of the projections, their great number and their spatial dispersion make the viewing of bursting difficult with the usual techniques of observation (video or photography). The first question to answer is whether the rupture occurs at a single point in the bubble cap or simultaneously at several points on the surface of the film.

According to some authors,[10,12] the surface of the film is covered by unstable capillary waves, which leads to severe

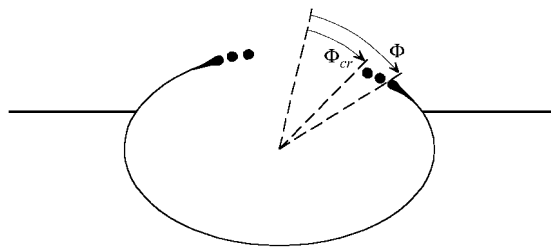

Film drops are formed by detachment if and only if $\Phi \geq \Phi_{cr}$ with $\Phi_{cr}=31.3°$

Fig. 4. Spiel's[15] criterion for film drop formation.

thickness fluctuations within the bubble cap. Therefore, the structure of the film would look like a network of thick ligaments linked by fine membranes, which may break up nearly simultaneously, giving birth to film drops.

For others,[1,7,15] the bubble cap begins to disintegrate by rupturing at a single point; the hole so formed rapidly widens, its edge forms a toroidal roll in which the film matter accumulates. Film droplets are detached when the surface tension forces become insufficient to prevent the roll from tearing loose. Spiel[15] showed that this condition leads, after calculations, to a geometrical criterion of film drop formation. This criterion applies to the angle $\Phi$ through which the roll has advanced since the hole opened (**Fig. 4**), and it is independent of the nature of the liquid and the gas.

The disintegration of the roll and thus the outbreak of the film drops occur when $\Phi$ reaches the critical value of 31.3°. This theory is valid only for a rupture at a single point in the bubble cap. By using this criterion and the calculation of Unger et al.[16] on the size and shape of a floating bubble, it is possible to determine a critical bubble size under which any bubble cannot launch any film droplet. In the case of water, this critical size is 2.4 mm and has been verified experimentally by Spiel.[15] In the case of liquid steel, the calculations produce a critical size between 3 and 4.5 mm depending on the composition of the steel (higher or lower concentration in sulfur and oxygen).[1]

In other respects, many authors[12,13,15,17] studied the number and size of film drops as a function of the bubble size. The number is proportional to the film area. The size distribution is large: from 0.3 to 500 $\mu$m.

After the disruption of the bubble cap, the cavity remaining at the liquid surface closes up, creating an upward Rayleigh jet that is unstable and can break up into droplets usually called jet drops (Fig. 3(c)). The number of jet drops never exceeds ten and decreases when the bubble size increases. Their size has been found to range between 0.1 and 0.18 times the diameter of their parent bubble for air–water system.[7,14] For a bubble diameter greater than 1 mm, the formation of jet drops can therefore be observed by photography or video, contrary to the formation of film drops.

## 3. Experimental Device

For studying dust emission from bubble burst in liquid steel, we set up an original experimental device (**Fig. 5**) using a vacuum induction furnace (Leybold). This furnace is operated at atmospheric pressure under an argon atmos-phere. The aims of the experiments are to clarify the way the bubbles burst at a liquid steel bath surface and to quantify the resulting emissions, i.e. film drops and jet drops.

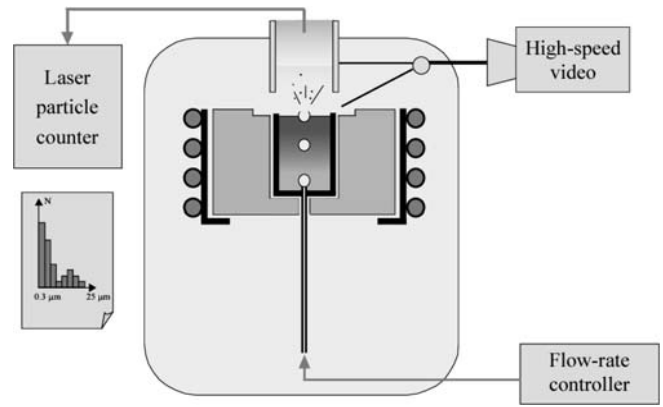

Fig. 5. Schematic diagram of the experimental device.

The steel charge (750 g of a commercial steel grade XC38) is melted in an alumina crucible, fitted in a graphite susceptor. This configuration reduces electromagnetic convection in the metal bath. The gas injection device consists of an alumina tube, fed with gas by a stainless steel tube, which is connected, outside the furnace, to a mass flow controller and the argon cylinder. In order to change the bubble size, we use three different sizes of alumina capillaries (outer diameter: 3, 1.2 or 0.5 mm). Moreover, for a given capillary, the gas flowrate can be modified (between 1 cm$^3$ min$^{-1}$ and 15 cm$^3$ min$^{-1}$) as well as the pressure drop, which enables the variation of the bubble size in a wide range. Currently, bubbles with diameter between 4.5 and 12 mm can be produced by this device.

The bursting of bubbles at the surface and the formation of the jet drops are observed by means of a high-speed video camera (Kodak Motion Corder) which enables to film the bath surface up to a rate of 10 000 frames per second. Actually, good-quality images could not be obtained at such a rate because increasing the shooting frequency entails a reduction in image resolution. We therefore selected rates of 5 000 frames s$^{-1}$ to record the film break and of 1 000 frames s$^{-1}$ to determine the frequency of emergence of the gas bubbles at the surface. The later frequency is equal to the frequency of the bubble formation at the capillary mouth and thus permits the calculation of the bubble size (equivalent-volume diameter) knowing the gas flowrate.

Lastly, to study the film drops, the aerosol formed is exhausted through a rack-mounted tube. The airborne particles are then characterized in number and size with a laser particle counter (Malvern ACP 300, range 0.3 to 25 $\mu$m). The exhaust flowrate is $4.39 \times 10^{-4}$ Nm$^3$ s$^{-1}$, which corresponds to a gas velocity of 0.4 m s$^{-1}$ at 500 K (typical gas temperature above the bath). According to the Stokes law, it enables to carry up particles up to 60 $\mu$m in diameter, a size which is greater than that of the largest particles found in EAF dust. Almost all of the particles analysed by the counter can be considered as coming from the steel bath. Indeed, it is possible to remove most of the parasitic particles present in the furnace atmosphere by sweeping it with filtered gas. At the beginning of each experiment, the fur-nace is pumped out and then fed with filtered argon. After one hour of sweeping, there remains in the furnace less than 100 particles with a diameter greater than 0.3 µm for 28.3 L of gas and none of these particles have a diameter above 1 µm. These experimental precautions enable a sufficient cleanness of the furnace to ensure an accurate determination of the emissions coming from the steel bath.

## 4. Results

### 4.1. Film Break

The high-speed video recording clearly provides evidence that the bubble cap ruptures from one point, whatever the bubble size. To our knowledge, these are the first images of bubble burst at the surface of a steel bath. **Figures 6** and **7** show two film break sequences. In these figures, the liquid steel bath appears in gray, the bubble in black and the hole in the bubble cap in white. The hole, first visible in the second frame of both sequences, can appear at any position on the cap, just as easily at the center or at the edge. Then, the hole widens until the film has completely disappeared, generally in less than 1 ms. No film drop projections could be observed, due to their small size and the limited image resolution (one pixel corresponds to 180–200 µm).

The single point break result is important because it allows to retain Spiel's criterion for the detachment of film drops in the case of liquid steel.

### 4.2. Jet Drops

The frames obtained after the disintegration of the bubble cap confirmed the formation of an upward liquid jet from which some visible droplets can be ejected.

We wished to determine the size of jet drops in order to determine to which extent these contribute to dust formation. The results obtained from 8 experiments reveal that the jet drop size represents 10 to 18% of the parent-bubble size, which is in agreement with the data from literature for the air-water system. **Figure 8** shows four jet drop size distributions obtained from bubbles of diameter 7.1, 7.4, 8.2 and 8.7 mm. The median jet drop diameter corresponds to 12, 15.8, 16.9 and 14.6%, respectively, of the parent bubble diameter. Such drops are too large to be exhausted by the experimental fume extraction system. This is confirmed by the video sequences that show that all the jet drops observed fall back down into the bath or around the crucible.

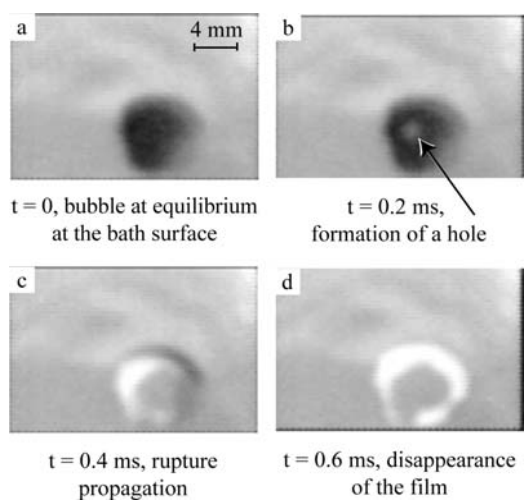

**Fig. 6.** Burst of a 7-mm diameter bubble at the surface of liquid steel (consecutive frames taken from a video sequence at 5 000 frames per second).

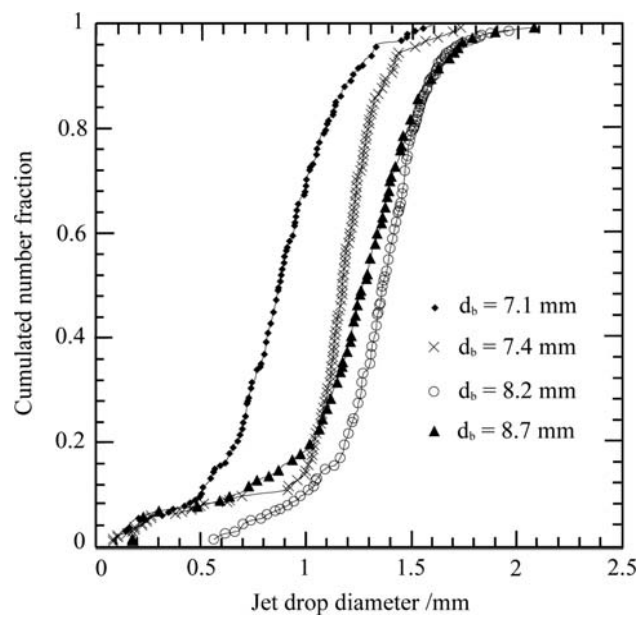

**Fig. 8.** Examples of size distribution of the jet drops collected after the burst of bubbles of diameters 7.1, 7.4, 8.2 and 8.7 mm.

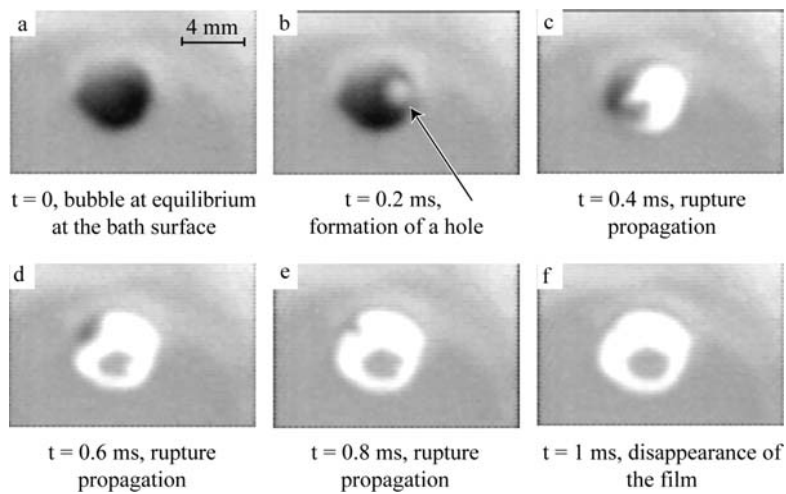

**Fig. 7.** Burst of a 6-mm diameter bubble at the surface of liquid steel (consecutive frames taken from a video sequence at 5 000 frames per second).

Table 1. Probability of jet drop projection *versus* parent bubble size.

| Average diameter of the bubbles observed | Number of bubbles observed | Number of jet drops projected | Number of jet drops per bubble ($N_{drops}$) |
|---|---|---|---|
| 4.9 mm | 106 | 265 | 2.5 |
| 6.0 mm | 110 | 143 | 1.3 |
| 7.1 mm | 83 | 58 | 0.7 |
| 8.7 mm | 68 | 19 | 0.28 |
| 10.2 mm | 58 | 7 | 0.12 |
| 12.4 mm | 33 | 1 | 0.03 |

We also determined the number of jet drops for different sizes of bubbles from the analysis of the video frames of bubble burst. The results (**Table 1**) are consistent with those presented in the literature: the probability of jet drop formation, and thus the number of jet drops per bubble, increases as the size of the parent bubble decreases. From these results, an exponential law similar to those proposed by Blanchard[8] or Wu[17] in the case of air–water systems can be obtained:

$$N_{drops} = 43.4 \exp(-0.58 d_B) \quad \quad (1)$$

where $d_B$ is the bubble diameter in mm and $N_{drops}$ the number of jet drops per bubble.

### 4.3. Film Drops

To quantify the film drop emission, we analyzed aerosols from experiments with and without bubbling, using the particle counter. Without bubbling, the particles detected come from vaporization of the steel charge, clearly visible in the form of fumes inside the furnace. Most of these particles are in a size range from 0.7 to 2 μm. Under the same experimental conditions, but with argon bubbling, larger particles appear in the aerosol size distribution, while the finer fractions decrease, the later effect being possibly due to agglomeration phenomena in the extraction tube. **Figures 9** and **10** present the results obtained for two bubble sizes for one minute of sampling. Due to the variation of the bubbling rate $Q_B$ with the bubble size, Fig. 9 corresponds to the burst of 162 bubbles whereas Fig. 10 corresponds to the burst of 106 ones.

From these results, it is possible to calculate the mass of the projections (right axis in Figs. 9 and 10) making a few simplifying assumptions:
- particle deposition on tube walls is neglegible,
- the average particle size of each class given by the counter corresponds to the arithmetic average of the class limits. For example, we consider every particle contained in the class 1–2 μm to be 1.5 μm in diameter,
- the density of every particle is 7 000 kg m$^{-3}$.

We obtain the mass of emissions due to the bubble-burst phenomenon only, by difference between the results of the experiments with- and without bubbling. According to the results of the jet drops study, these emissions are only composed of film drops. Thus, for the 6.1-mm bubbles, the mass of the projections ($M_p$ in **Table 2**) is, on average, 231 μg for 1-min sampling, compared with 578 μg in the case of the 7.9-mm bubbles. For an easier and more realistic comparison, these results have been related to one bubble burst ($M_B$ in Table 2) and to the volume of injected gas ($M_g$ in Table 2), which is a unit of measure used by steelmakers. The results (Table 2) clearly show that the amount of film drops coming from bubble burst at the surface of the steel bath decreases with the bubble size.

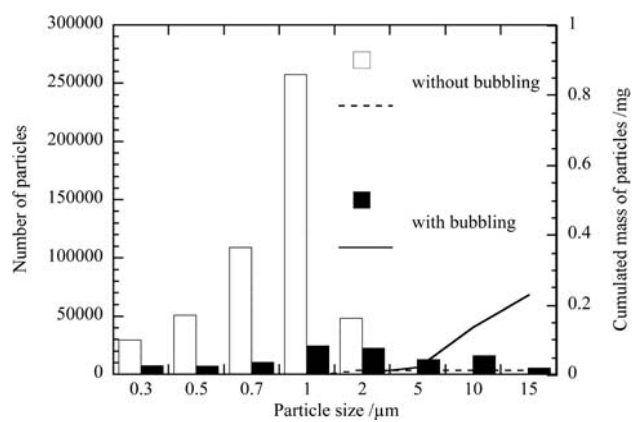

Fig. 9. Particle size distributions and aerosol masses, for a 1-min sampling with 6.1-mm bubbles.

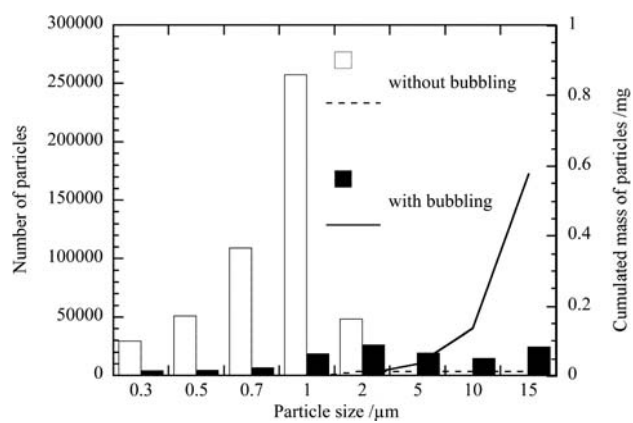

Fig. 10. Particle size distributions and aerosol masses, for a 1-min sampling with 7.9-mm bubbles.

Table 2. Mass of film drops *versus* parent bubble size.

| $d_B$ (bubble diameter) | $Q_B$ (bubbling rate) | $M_p$ (mass of projections for 1-min sampling) | $M_B$ (mass of projections for one bubble burst) | $M_g$ (mass of projections for 1 m$^3$ of injected gas) |
|---|---|---|---|---|
| 6.1 mm | 162 bubbles/min | 231 μg/min | 0,138 μg/bubble | 0,012 kg/m$^3$ |
| 7.9 mm | 106 bubbles/min | 578 μg/min | 0,532 μg/bubble | 0,021 kg/m$^3$ |

### 5. Discussion

In EAF, 60% of dust comes from projections of liquid metal and slag (from Fig. 2). These projections from the bath can be attributed to the CO-bubble burst. The mechanisms involved in the formation of those projections in EAF can be a little different from those found in our experiment, particularly owing to the presence of a slag, either foaming or not. Nevertheless, the results we have obtained provide useful information for the understanding and the quantification of dust formation in EAF.

Jet drops come from the disintegration of the upward jet created after the removal of the bubble cap. Their number increases when the bubble size decreases, and their size represents 10 to 18% of the parent bubble size. The size of CO-bubbles formed in EAF remains poorly known. These bubbles result from the decarburization reaction, which involves nucleation and growth mechanisms. Nucleation can occur at various sites (furnace walls, inclusions, slag droplets...). The rate of growth depends on local conditions, such as local oxygen activities. Although the exact size distribution of CO-bubbles in EAF is unknown, it is yet possible to assess indirectly the size range. For example, the stirring phenomena in the bath are correctly modelled assuming bubble diameters varying between 1 and 2 cm.[1] Besides, foaming slag samples show after solidification the presence of bubbles of a few millimeters in diameter.[11] To sum up, CO-bubble size is probably between 2 and 20 mm. According to our results, this kind of bubbles would give rise to jet drops whose size varies from 0.2 mm to almost 4 mm, which is much higher than the size of particles found in EAF dust samples.[1] As observed in our laboratory furnace, jet drops are not exhausted by the fume extraction system and are likely to fall back down into the steel bath. Jet drops can thus hardly participate in dust formation from bubble burst in EAF.

Film drops are emitted during the disintegration of the liquid cap which covers the bubble at the surface of the bath. Their size distributions are very close to that of the particles contained in EAF dust[1] (see Figs. 9 and 10). The results obtained for two sizes of bubbles (6.1 mm and 7.9 mm) lead to amounts of film drops of 0.012 kg per $m^3$ of gas injected for the smaller bubbles and 0.021 kg m$^{-3}$ for the bigger ones. EAF produces between 15 and 25 kg of dust per ton of steel. 60% of the dust comes from the bursting of CO bubbles. Knowing that 544 $m^3$ of CO are evolved for making 1 t of steel, the corresponding amount of projections then ranges between 0.016 and 0.028 kg m$^{-3}$. These figures are close to those derived from our laboratory experiments. Associated with the conclusion of the jet drop size study, they show that, when a bubble bursts, it is mostly the film drops that contribute to dust formation. These first results also reveal a significant decrease in the amount of film drops resulting from bubble burst when the bubble size decreases. Unfortunately, our experimental device could not yet produce bubbles small enough to check, in the case of liquid steel, the existence of a critical bubble size (4 mm) derived from Spiel's criterion.[15]

A theoretical solution for dramatically reducing the amount of dust produced in EAF can be derived from our work: decreasing the CO-bubble size, ideally between 1 and 4 mm. The latter bound prevents the film drop formation and the former one avoids the emission of jet drops small enough to be carried up. Such an objective may be difficult to reach since the CO-bubble formation is a rather spontaneous process. Nevertheless, a possible method would be to better control the decarburization reaction, for example by favoring nucleation at the expense of growth.

## 6. Conclusions

The study of dust formation in an electric arc furnace has shown that the projections resulting from the CO-bubble burst at the surface of the liquid steel bath represent the main mechanism for dust emission. We have therefore designed an experimental device to observe the gas bubble burst at the surface of liquid steel by means of high-speed video, and to quantify the resulting projections with a laser particle counter. The phenomena involved are similar to those taking place in the case of an air bubble bursting at water surface: the rupture of the liquid cap of the bubble starts from a single hole and results in the formation of an aerosol of very small droplets, called film drops; after the film completely vanished, emerges a liquid jet, from which are detached a few drops, the jet drops, too big to be captured by the gas extraction system. We have also shown that the amount of film drops decreases with the size of the parent bubble. Theoretically, it would be possible to produce no film drop, and therefore to reduce drastically the dust emission, if all the bubbles formed were smaller than 4 mm in diameter.

The continuation of the present piece of work will consist in studying the influence of surfactant elements, such as S or O, and the presence of a slag covering the liquid steel, on the bubble burst.

**Acknowledgements**

We thank O. Devisme for his help in the experimental part of the work. The present study has been supported by CNRS (French National Scientific Research Center), the Lorraine region and IRSID (Arcelor group).